\documentclass[conference]{IEEEtran}
\IEEEoverridecommandlockouts

\usepackage{cite}
\usepackage{amsmath,amssymb,amsfonts}
\usepackage{algorithmic}
\usepackage{graphicx}
\usepackage{textcomp}
\usepackage{xcolor}
\usepackage{booktabs}
\usepackage{multirow}
\usepackage{hyperref}
\usepackage{tabularx}
\usepackage{eso-pic}   

\def\BibTeX{{\rm B\kern-.05em{\sc i\kern-.025em b}\kern-.08em
    T\kern-.1667em\lower.7ex\hbox{E}\kern-.125emX}}

\begin{document}


\title{AI-driven Intent-Based Networking Approach for Self-configuration of Next Generation Networks}

\author{
\IEEEauthorblockN{
Md. Kamrul Hossain\IEEEauthorrefmark{1},
Walid Aljoby\IEEEauthorrefmark{2}
}
\IEEEauthorblockA{
\IEEEauthorrefmark{1}\IEEEauthorrefmark{2}Information and Computer Science Department, King Fahd University of Petroleum and Minerals, Dhahran 31261, Saudi Arabia\\
\IEEEauthorrefmark{2}IRC for Intelligent Secure Systems, King Fahd University of Petroleum and Minerals, Dhahran 31261, Saudi Arabia
}
\IEEEauthorblockA{(e-mail: g202215400@kfupm.edu.sa, waleed.gobi@kfupm.edu.sa)
}
}

\maketitle

\AddToShipoutPictureFG*{%
  \AtPageLowerLeft{%
    \raisebox{40pt}{%
      \makebox[\paperwidth]{%
        \centering\footnotesize
        © 2026 IEEE. Accepted for presentation in IEEE/IFIP NOMS 2026.
      }%
    }%
  }%
}

\begin{abstract}
Intent-Based Networking (IBN) aims to simplify operating heterogeneous infrastructures by translating high-level intents into enforceable policies and assuring compliance. However, dependable automation remains difficult because (i) realizing intents from ambiguous natural language into controller-ready policies is brittle and prone to conflicts and unintended side effects, and (ii) assurance is often reactive and struggles in multi-intent settings where faults create cascading symptoms and ambiguous telemetry. This paper proposes an end-to-end closed-loop IBN pipeline that uses large language models with structured validation for natural language to policy realization and conflict-aware activation, and reformulates assurance as proactive multi-intent failure prediction with root-cause disambiguation. The expected outcome is operator-trustworthy automation that provides actionable early warnings, interpretable explanations, and measurable lead time for remediation.
\end{abstract}


\vspace{-0.3cm}
\section{Introduction and Motivation}
Software-Defined Networking (SDN) decouples control and forwarding to enable centralized programmability and is a key enabler for automating modern networks. Large-scale deployments such as Google's B4 and Microsoft's SWAN demonstrate its practicality and scalability~\cite{hong2018b4}. Yet, operators still must translate business goals into low-level controller rules and configurations, which is a time-consuming and error-prone process. Even with mature controller ecosystems and tooling, translation and validation remain largely manual. As networks evolve toward 6G and cloud-native infrastructures, they must meet strict service requirements while also supporting microservices-based east–west traffic, multi-tenancy, and zero-trust segmentation. As a result, configuring each device individually is becoming increasingly complex and impractical~\cite{marinova2024intelligent, 10.1145/3232755.3234555}. 

Intent-Based Networking (IBN) bridges the gap between \emph{what} operators want and \emph{how} the network is configured~\cite{9925251}, with growing industry adoption. Following RFC~9315~\cite{rfc9315}, an \emph{intent} is a declarative statement of goals and expected outcomes, independent of realization mechanisms. Realizing intents requires compiling them into policy artifacts and enforcing them via controllers and orchestrators, including SDN fabrics and Open RAN SMO/RIC-based stacks~\cite{d2023orchestran}. Although IBN aims to automate translation, activation, and closed-loop validation, many systems still require structured intent formats (e.g., JSON/XML/YAML/NSD)~\cite{9925251} and bind them to specific data models and controller semantics, limiting usability and increasing misconfiguration risk in heterogeneous ecosystems.

A second barrier is \emph{intent assurance} which is continuously verifying that operational state satisfies intended goals~\cite{10575429}. Assurance is central to closed-loop automation in 5G/6G, reflected by the 3GPP NWDAF analytics function~\cite{9824403}. Intent violations are often preceded by \emph{intent drift}~\cite{rfc9315} which is first visible as weak but persistent KPI changes~\cite{10575429, 9925251}. Many solutions remain reactive~\cite{10575429, 11073595}, and the problem is harder in \emph{multi-intent} settings where interacting intents produce cascading symptoms and ambiguous co-drifting KPIs~\cite{11073595, 10575429, 9615580}. For example, in cloud-native edge applications, resource contention affecting one service can degrade KPIs elsewhere, creating symptomatic \emph{victims} that are hard to distinguish from the root-cause intent~\cite{10.1145/3501297, 10.1145/3580305.3599934}.

This research targets dependable end-to-end IBN via two complementary thrusts. First, we investigate how large language models (LLMs) can enable \emph{natural language to policy} realization with bounded failure modes, using schema-constrained policy intermediate representation (IR) generation, context examples, and conflict-aware activation across heterogeneous controllers. While recent efforts explore LLMs for networking/IBN~\cite{liu2024large}, most emphasize translation and do not systematically address conflict detection, activation robustness, and closed-loop assurance across realistic controller stacks~\cite{10574890, han2025network, wang2024netconfeval, tu2025intent, li2024preconfig}. Second, we reformulate assurance as \emph{proactive failure prediction} under multi-intent ambiguity. This includes learning early KPI precursors, disambiguating root-cause intents from co-drift victims, and providing operator-facing outputs such as attribution and lead-time (early warning) estimates via multi-horizon modeling.

Collectively, the goal is an operationally viable closed-loop IBN pipeline that (i) translates natural-language intents into verifiable, controller-ready policy artifacts, (ii) activates them safely through conflict-aware deployment, and (iii) assures compliance proactively in multi-intent environments with interpretable, actionable outputs.

\vspace{-0.1cm}
\section{Related Work}
IBN automates network operation through an \emph{intent lifecycle} that includes translation, activation, and assurance~\cite{9925251}. While substantial progress has been made in individual stages, dependable \emph{end-to-end} automation remains difficult in practice, particularly under heterogeneous controller ecosystems and multi-intent operational dynamics.

\textbf{Translation and realization:}
Earlier systems used rule-based NLP and programmatic logic to parse intents and generate configurations~\cite{jacobs2019deploying,9925251}, but natural-language variability and context dependence limit their robustness. Recent work leverages LLMs for intent translation and configuration generation~\cite{11073595, han2025network,mekrache2024llm,wang2024netconfeval,10574890}. For example, NetConfEval benchmarks LLMs for mapping specifications to structured rules~\cite{wang2024netconfeval}. However, the studies rely mostly on closed-source models, evaluate limited model/dataset combinations, and often stop at abstract outputs rather than controller-ready artifacts, leaving open questions about robustness across practical controller schemas.

\textbf{Activation and conflict handling:}
Deploying intents safely requires preventing conflicts among overlapping policies. Prior SDN/IBN work categorizes conflicts and provides detection methods, supported by verification and analysis tools such as VeriFlow, NetPlumber, and Batfish~\cite{8453007}. Additional approaches use logical evaluation, graph reasoning, and transactional mechanisms~\cite{zhang2021conflict,9925251}. Yet, these systems often provide limited actionable guidance on \emph{why} conflicts occur and how to resolve them, and LLM-based conflict handling has so far been narrow in scope~\cite{wang2024netconfeval}. Controller-aware activation that is both verifiable and operationally actionable remains underexplored.

\textbf{Assurance and drift:}
Intent assurance verifies that operational behavior continues to satisfy intents~\cite{10575429}. Beyond static checks and flow-level invariants~\cite{11293797, 8453007,9925251}, many approaches monitor KPIs and detect violations after degradation becomes visible. Predictive approaches forecast individual KPIs (e.g., LSTM-based forecasting with thresholding)~\cite{9615580} or classify near-term events (e.g., congestion prediction). Besides, unsupervised drift detection can help when labels are scarce~\cite{10770652}, but may be prone to false alarms and is not optimized to learn subtle failure precursors. Recent methods treat drift as KPI deviation from targets and use LLMs to generate corrective actions~\cite{10575429,11073595}, but remain largely reactive and typically do not estimate urgency (e.g., time-to-failure). Critically, existing approaches also struggle with \emph{multi-intent ambiguity}, where one fault yields cascading symptoms across multiple intents~\cite{10.1145/3580305.3599934}.

\textbf{Positioning:}
Thus, existing work either emphasizes partial lifecycle automation (often translation)~\cite{han2025network,mekrache2024llm,wang2024netconfeval,tu2025intent,10574890,li2024preconfig} or provides assurance methods that are reactive or single-intent oriented~\cite{hossain2026leaddriftrealtimeexplainableintent, 10575429,11073595,10770652,9615580} and can be challenging to scale. This thesis addresses these gaps by pursuing an end-to-end, controller-aware IBN pipeline that combines verifiable LLM-assisted realization (translation and conflict-aware activation) with proactive \emph{multi-intent} failure prediction and root-cause disambiguation, extending predictive outage modeling ideas~\cite{Basikolo2023TowardsZD,9557387} to the specific challenges of IBN assurance.

\section{Proposed Approach}

\begin{figure*}[t!]
    \centering
    \includegraphics[width=\linewidth]{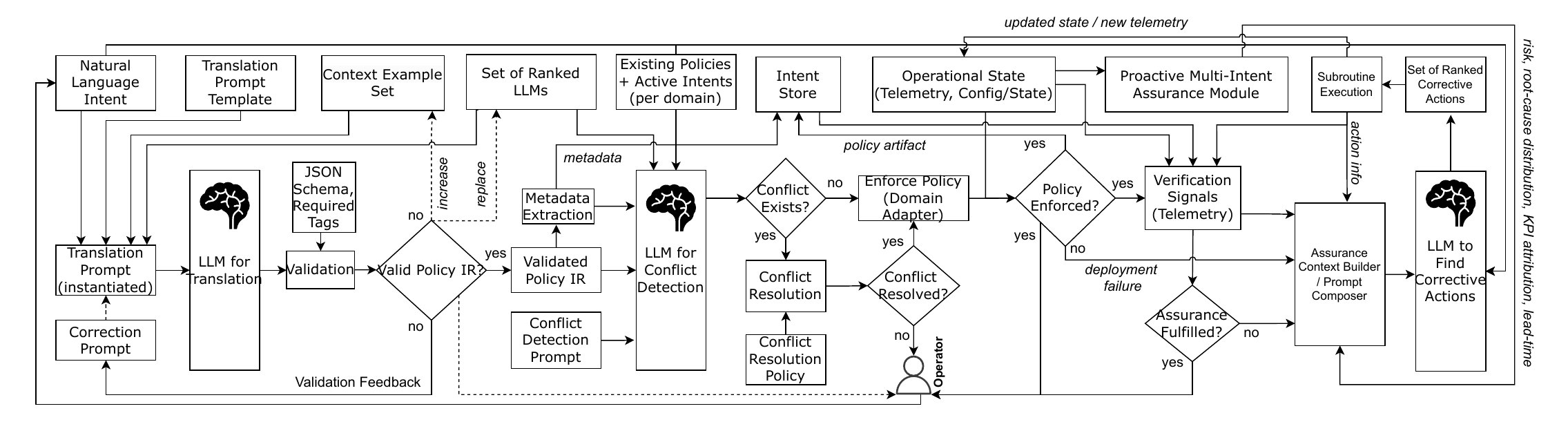}
    \caption{Closed-loop thesis framework: LLM-assisted realization (translate+activate) coupled with proactive multi-intent assurance and ranked remediation.}
    \label{fig:mild_framework}
\end{figure*}

Fig.~\ref{fig:mild_framework} illustrates a holistic architecture of our proposed approach. It instantiates the classical IBN loop---\emph{translate, activate, assure}---as a single pipeline that explicitly couples (i) verifiable intent realization, (ii) controller-aware conflict-safe activation, and (iii) proactive multi-intent assurance with actionable remediation. At a high level, natural-language intents are compiled into a schema-constrained policy IR, activated through domain adapters with enforcement feedback, and continuously assured using telemetry and observed state where assurance outputs (risk, likely root cause, attribution, lead time) drive ranked corrective actions that close the loop.

We elaborate these stages in the following subsections, starting with verifiable LLM-based realization, then conflict-aware activation and enforcement checks, and finally, proactive multi-intent assurance and ranked remediation.

\subsection{LLM-Assisted Intent Realization with Verifiability}
Given a natural-language intent, the framework instantiates a translation prompt from a prompt template and context examples, then invokes an LLM to produce a \emph{structured policy IR} (e.g., JSON/YANG-like). To bound failure modes, the output is validated against a required schema/tags where invalid outputs trigger a correction prompt and re-generation until a valid policy IR is produced. In parallel, metadata (e.g., scope, constraints, priority) is extracted from the validated policy IR to support downstream conflict checks and assurance context building.

\subsection{Conflict-Aware Activation and Enforcement Checks}
Before deployment, the framework performs conflict detection by combining the validated policy IR, extracted metadata, and the set of existing policies/active intents (per domain). A ranked/selected LLM (router-style selection) can be used to specialize conflict analysis across domains and intent types (e.g., mixture-of-experts style selection \cite{chen2022towards}). If conflicts are detected, the framework applies a conflict-resolution policy and iterates until the conflict is resolved, and unresolved cases are escalated to the operator.
Once cleared, the policy artifact is enforced through a domain adapter, followed by an explicit \emph{policy enforcement} check. Importantly, enforcement failures (e.g., deployment/runtime errors) are directly fed into the assurance context builder so corrective actions can be generated even when realization does not complete successfully.

\subsection{Proactive Multi-Intent Assurance and Remediation}
The assurance stage continuously aggregates operational state (telemetry and observed configuration/state) and feeds it to a proactive multi-intent assurance module, e.g., a fixed-horizon predictor with intent-level root-cause disambiguation under co-drift~\cite{hossain2026mildmultiintentlearningdisambiguation}. This module outputs early-warning signals, including per-intent risk and root-cause indications under multi-intent ambiguity, as well as KPI-level attribution and lead-time estimates. In addition, verification signals derived from telemetry provide near-real-time evidence of whether the current assurance objectives are satisfied.

To drive remediation in a consistent way, the framework uses an \emph{assurance context builder / prompt composer} that fuses: (i) verification signals, (ii) proactive module's outputs (risk/root-cause/attribution/lead-time), and (iii) intent/policy context from the intent store. This composed assurance prompt is passed to an LLM that produces a ranked set of corrective actions. The selected actions are executed as subroutines, after which the updated state and new telemetry close the loop by feeding back into the operational state, proactive assurance module, and enforcement/assurance checks.

\subsection{Telemetry Signals for Multi-Intent Assurance}
We aim to leverage multiple representative application KPIs that can be used as inputs to the proactive assurance and verification stages such as CPU, RAM and storage utilization percentage, Service network throughput, Service reliability/availability index, End-to-end API request latency, Ingestion/queue backlog length and Analytics/processing throughput.

\vspace{-0.5cm}
\section{Challenges and Open Questions}
While the proposed framework (Fig.~\ref{fig:mild_framework}) outlines an end-to-end pipeline for verifiable intent realization and proactive multi-intent assurance, several challenges must be addressed before reliable deployment at scale, especially in heterogeneous and 5G/6G environments.

\textbf{Federation and 5G/6G multi-layer control:}
Moving beyond single-domain SDN requires operating across federated domains and multi-layer control planes (core, transport, RAN), where controllers expose different abstractions, policies, and trust boundaries~\cite{marinova2024intelligent}. A practical direction is to maintain a controller-agnostic policy IR at the global layer, while using domain adapters to compile/validate domain-specific artifacts under explicit \emph{policy contracts} that govern what telemetry and state can be shared~\cite{10.1145/3232755.3234555}. Key open questions are: (i) what minimal shared schema (or capability description) is sufficient for cross-domain conflict detection; (ii) how to provide safe activation guarantees when domains can reject, delay, or partially apply policies; and (iii) how to validate end-to-end compliance for multi-layer intents (e.g., slice-like intents) when layers differ in observability and timescales.

\textbf{Data, labels, and generalization for proactive assurance:}
Proactive multi-intent assurance depends on learning subtle precursors of failure from KPI streams~\cite{11073595}, yet labeled multi-intent co-drift incidents are scarce and telemetry is often noisy, missing, delayed, or biased by sampling. We therefore pursue \emph{hybrid evaluation} using controlled benchmark generation for specific co-drift patterns, complemented with testbed traces and partially labeled operational logs when available. Open questions include: (i) which failure/co-drift patterns best approximate realistic incidents, (ii) how to quantify domain shift between synthetic and real telemetry, and (iii) how to calibrate risk estimates so lead-time predictions remain trustworthy across environments.

\textbf{Safety controls for LLM-assisted realization:}
Natural-language intent expression introduces risks of incorrect or unsafe outputs (e.g., malformed artifacts or over-permissive security policies). Beyond schema validation and conflict-aware activation, deployment requires stronger controls such as constrained decoding to a typed IR, static checks (e.g., invariants and policy overlap), and staged activation (canary and rollback)~\cite{11293797, wang2024netconfeval}. A central open question is the \emph{safety contract}, namely which properties can be guaranteed pre-deployment (e.g., syntactic validity and non-contradiction with existing intents) versus those requiring runtime evidence (telemetry-based verification and targeted active tests). Another practical concern is \emph{prompt grounding}. Prompts should reflect the current network state without exposing sensitive details, so the assurance context builder should supply only minimal summaries (e.g., topology sketches, intent inventory, critical KPIs)..

\textbf{Repair and self-reconfiguration:}
When drift, faults, or intent violations are observed, the loop must trigger \emph{safe repair} actions such as re-configuration, re-routing, scaling, or policy re-optimization without violating coexisting intents. Practical directions include policy-aware remediation workflows, staged execution with post-change verification, and human-in-the-loop approval for high-impact changes. Open questions include how to rank actions by expected impact/risk under uncertainty and how to prevent control-loop instability in multi-intent settings.

\textbf{Latency, scalability, and actionable outputs:}
End-to-end automation must meet operational time budgets where LLM inference, conflict analysis, and predictive assurance can add latency under frequent churn or large intent volumes~\cite{tu2025intent, 9925251}. A pragmatic approach is to separate tasks by criticality and timescale where lightweight validation/conflict filtering runs inline, while heavier reasoning (explanations and deep predictive models) executes asynchronously with rate limits and caching. Finally, predictions must be actionable. Beyond accuracy and lead time, evaluation should capture alert fatigue, root-cause disambiguation quality, and the safety/impact of recommended actions under multi-intent interactions. Open questions include how to partition intelligence across edge vs.\ central placement, and which explanation modalities (KPI attribution, causal graphs, counterfactuals) best support operator trust and remediation.


\vspace{-0.3cm}
\section{Conclusion}
This doctoral research investigates how to make intent-based networking more practical and dependable by unifying two capabilities in a single closed loop: (i) verifiable, LLM-assisted intent realization that translates natural-language intents into controller-ready policy artifacts with conflict-aware activation; and (ii) proactive, multi-intent assurance that anticipates failures early and disambiguates the likely root cause from co-drifting symptoms. To date, we have implemented and evaluated key realization components in SDN settings (structured policy generation/validation and controller-aware activation), and we have validated the proactive assurance direction on controlled multi-intent scenarios, showing promising lead time and improved root-cause separation. However, two open questions remain in our research: \\
\textbf{Extending to 5G/6G management:} What is a realistic, well-scoped path to extend the same pipeline beyond SDN into 5G/6G (e.g., slice-like/service intents spanning core/transport/RAN), and which initial intent types and enforcement points provide the most defensible evaluation?\\
\textbf{Validation with real telemetry:} What is the most practical strategy to evaluate proactive multi-intent assurance on real data---adapting existing datasets via augmentation/transfer learning, instrumenting a controlled testbed to collect KPI traces with incident labels, or combining both---and what labeling/evaluation protocol best supports credible root-cause and lead-time claims?

\bibliographystyle{IEEEtran}
\vspace{-0.5cm}
\bibliography{references}

\end{document}